\documentclass[twocolumn,aps,prl,letter,nofootinbib]{revtex4-1}
\pdfoutput=1

\usepackage{amsmath}
\usepackage{amsfonts}
\usepackage{graphicx}
\usepackage{amssymb}
\usepackage{mathtools}
\usepackage{mathrsfs}
\usepackage{titlesec}
\usepackage{hyperref}
\usepackage[nottoc,numbib]{tocbibind}
\usepackage[titletoc]{appendix}
\usepackage{adjustbox}
\usepackage{blkarray}
\usepackage{tabularx}
\usepackage{cleveref}
\usepackage{enumitem}
\usepackage{tensor}
\usepackage{caption}

\begin{document}
\title{Scalar Hairy Black Holes in Four Dimensions are Unstable}

\author{Bogdan Ganchev}
\email{bvg25@cam.ac.uk}

\author{Jorge~E.~Santos}
\email{jss55@cam.ac.uk}
\affiliation{Department of Applied Mathematics and Theoretical Physics, University of Cambridge, Wilberforce Road, Cambridge CB3 0WA, UK}

\begin{abstract}
\noindent{We present a numerical analysis of the stability properties of the black holes with scalar hair constructed by Herdeiro and Radu. We prove the existence of a novel gauge where the scalar field perturbations decouple from the metric perturbations, and analyse the resulting quasinormal mode spectrum. We find unstable modes with characteristic growth rates which for uniformly small hair are almost identical to those of a massive scalar field on a fixed Kerr background.}
\end{abstract}

\maketitle

{\bf~Introduction --} With the advent of gravitational wave astronomy by the LIGO collaboration, now also joined by VIRGO, \cite{scientific2016tests,abbott2016binary,scientific2017gw170104,abbott2017gw170814,abbott2017gw170817}, the understanding of asymptotically flat black holes (BHs) in four spacetime dimensions has taken a novel central role in theoretical physics. The study of BHs can be broadly divided into two complementary categories: a) search for stationary solutions and their concomitant stability analysis and b) strong field dynamics. While the latter is an active topic of research, the former was thought to have been understood during the seventies \cite{carter1971axisymmetric,robinson1975uniqueness} and lead to the formulation of the so-called \emph{no-hair} theorems. These state that if $(\mathcal{M},g)$ is a stationary, axisymmetric, four-dimensional asymptotically flat vacuum spacetime that is suitably regular on and in the vicinity of a connected event horizon, then it is isometric to a member of the Kerr family \cite{kerr1963gravitational}.

However, the \emph{no-hair} theorems do possess assumptions not all of which are physically well motivated\footnote{Earlier attempts for violating the \emph{no-hair} theorems have been presented in \cite{gubser2005phase}. However, such solutions fail to satisfy the positivity of energy theorem \cite{schoen1979proof,schoen1981proof,horowitz1982gravitational,gibbons1983positive} due to the existence of negative regions in the potential, thus bypassing the assumptions of the uniqueness theorems - see \cite{hertog2006towards}.}. Most notably, they assume the existence of a stationary Killing vector field (KVF) which is not the horizon generator. In \cite{dias2011black}, the first example of a hairy black hole (HBH) violating this assumption was constructed with anti de Sitter (AdS) boundary conditions. These solutions are time dependent and are \emph{not} axisymmetric from the matter perspective, but the gravitational sector does preserve axisymmetry and stationarity. These solutions were generalised in \cite{dias2015black}, where purely gravitational BH solutions with a single KVF were constructed\footnote{In this instance, the metric itself has a single KVF only.}.

Three key ingredients for constructing scalar hair were identified in \cite{dias2011black}: 1) confined scalar field so that bound states exist, 2) presence of superradiant scattering and 3) existence of a single Killing vector field, which happens to coincide with the horizon generator. A few years later, Herdeiro and Radu noticed that such a construction could work in asymptotically flat spacetimes \cite{herdeiro2014kerr} if a complex massive scalar field is minimally coupled to gravity. The idea being that the confining nature of AdS is replaced by the presence of the mass term. Building on this generalisation to asymptotically flat spacetimes, BHs with Proca hair have been recently constructed in \cite{herdeiro2016kerr}. In all of these cases\footnote{The exception being the five-dimensional case studied in \cite{brihaye2014myers}.}, the HBHs branch from the onset of the superradiant instability \cite{starobinskii1973amplification,damour1976quantum,detweiler1980klein,zouros1979instabilities,dolan2007instability,yoshino2014gravitational,Brito:2014wla,brito2015superradiance} and extend into regions of moduli space where Kerr BHs do not exist.

It is then interesting to investigate whether these HBHs are themselves unstable, since their stability analysis could have important consequences for their observation. This looks like a daunting task with little chance of success, since no Teukolsky equation has been found for the system at hand. It would seem one would have to perturb the full Einstein-Klein-Gordon (EKG) system and thus solve a complicated set of coupled linear partial differential equations. What is worse is that, since the background scalar field exhibits explicit time dependence, it would seem unlikely that the concept of quasinormal mode could be useful, since the time dependence of the fields would not factorise\footnote{That is to say, no useful Laplace transform can be taken to study stability using St\"urm-Liouville type methods.}. In order to bypass these issues, we will prove the existence of a new gauge where the scalar field perturbations decouple from the metric perturbations. Furthermore, we investigate the issue of residual gauge freedom, showing that our main results cannot be gauged away, thus rendering them physical.

Our paper is organised as follows: we first reconstruct the solutions of \cite{herdeiro2014kerr} and recover their results, then in section 2 we perturb the equations of motion and prove the existence of a particular gauge where the matter sector plays a pivotal role, followed by a discussion of our results, with the final section 4 dedicated to conclusions.
\\
\indent{\bf~Einstein-Klein-Gordon system --} We start with Einstein-Hilbert gravity minimally coupled to a complex massive scalar field
\begin{align}
S=\int_{\mathcal{M}} \mathrm{d}^4x\sqrt{-g}\left(\frac{R}{16\,\pi\,G}-\nabla_a\psi^*\nabla^a\psi-\mu^2|\psi|^2\right)\,.
\end{align}
From here henceforth we will set $G=c=1$. The corresponding equations of motion are given by
\begin{subequations}
\begin{equation}
\frac{R_{ab}}{8\pi}=2\nabla_{(a}\psi^*\nabla_{b)}\psi+g_{ab}\mu^2\psi^*\psi\,,
\label{eq:EKG1}
\end{equation}
\begin{equation}
\Box\psi=\mu^2\psi\,.
\label{eq:EKG2}
\end{equation}
\label{eq:EKG}%
\end{subequations}
A well-know solution to the above system of PDEs is the Kerr family of BHs, where the scalar field $\psi$ vanishes identically and
\begin{multline}
\mathrm{d}s^2 = -\frac{\Delta}{\Sigma^2}\left(\mathrm{d}t-a \sin^2\theta \mathrm{d}\phi\right)^2+\frac{\sin^2\theta}{\Sigma^2}[a\,\mathrm{d}t-(r^2+a^2)\mathrm{d}\phi]^2
\\
+\Sigma^2\left(\mathrm{d}\theta^2+\frac{\mathrm{d}r^2}{\Delta}\right)\,,
\label{eq:kerr}
\end{multline}
with $\Delta = r^2+a^2-2\,M\,r$ and $\Sigma^2=r^2+a^2\cos^2\theta$. The BH event horizon is a null hypersurface with $r=r_+\equiv M+\sqrt{M^2-a^2}$, angular velocity $\Omega_K = a/(a^2+r_+^2)$ and temperature $T_K=(r_+^2-a^2)/[4\pi r_+(r_+^2+a^2)]$. The constant $M$ is the BH mass and $a$ parametrises its angular momentum via $J = M\,a$. The absence of naked singularities demands $|a|\leq M$ with the inequality saturating at extremality, when the Kerr BH event horizon becomes degenerate with $T_K=0$.

In \cite{herdeiro2014kerr,Chodosh:2015oma} it was shown that HBHs can coexist with Kerr BHs in certain regions of the moduli space. Their existence in the phase diagram of \eqref{eq:EKG} can be understood via a linearised analysis of scalar perturbations on a fixed Kerr background. In such a spacetime scalar perturbations can be studied by taking $\psi = \hat{\psi}(r,\theta) e^{-i\,\omega\,t+i\,m\,\phi}$, with $\omega$ the frequency we wish to determine and $m\in\mathbb{Z}$ an azimuthal quantum number. If $\mathrm{Im}(\omega)>0$, the system exhibits a linear mode instability. The resulting equation for $\hat{\psi}(r,\theta)$ is separable into two ODEs that couple via their respective eigenvalues: one equation along the angular direction $\theta$ and one along the radial direction $r$.

The presence of an ergoregion can be used to extract energy from the BH and source superradiant \emph{scattering} \cite{bardeen1972rotating,damour1976quantum,starobinskii1973amplification,bekenstein1973extraction,Shlapentokh-Rothman:2013ysa}, so long as $0<\omega\leq m\Omega_K$. Since the scalar field is massive, these waves can be trapped and thus source an instability. This is the so called superradiant instability first uncovered in the late seventies and early eighties by Zouros and Eardley \cite{zouros1979instabilities} and Detweiller \cite{detweiler1980klein}. From the onset of this instability, novel hairy BHs bifurcate \cite{dias2015black} with $\omega=m\Omega_K$, thus preserving a single Killing vector field $\Xi = \partial/\partial_t+\Omega_K \partial/\partial_{\phi}$ only. Since the scalar field is complex, it yields a stress energy tensor which is axisymmetric and stationary and thus preserve as many isometries as those possessed by the Kerr line element (\ref{eq:kerr}). These BHs were constructed at the nonlinear level in \cite{herdeiro2014kerr,Chodosh:2015oma} and shown to coexist with the Kerr BH for certain regions of the $(M,J)$ plane, thus violating the uniqueness of the Kerr family of solutions.

As mentioned earlier, in order to assess their linear stability, we want to know whether these BHs are also susceptible to superradiance. To this end we perturb them and solve the resulting equations numerically, whereby a suitable choice of gauge reduces the system of equations to a KG equation for the scalar perturbation in a fixed HBH background. We therefore first construct these BHs to a very high accuracy using the DeTurck method, which was first presented in \cite{headrick2010new} and recently reviewed in \cite{Dias:2015nua}.
\\
\indent{\bf~Hairy Black holes --} Employing the DeTurck method in order to create HBHs amounts to solving the following system of PDEs:
\begin{multline}
R_{ab}-\nabla_{\left(a\right.}\xi_{\left.b\right)}=8\pi \left[2\nabla_{(a}\psi^*\nabla_{b)}\psi+g_{ab}\mu^2\psi^*\psi\right],
\label{eq:EinsteinDeTurck}
\end{multline}
where $\xi^a=g^{bc}\left[\Gamma^{a}_{bc}(g)-\Gamma^{a}_{bc}(\mathfrak{g})\right]$ is the DeTurck vector and $\Gamma^a_{bc}(\mathfrak{g})$ is the Levi-Civita connection for a reference metric $\mathfrak{g}$. The only restriction on $\mathfrak{g}$ is that it obeys the same boundary conditions (in the Euclidean sense) as the metric we wish to find.

Recall that we wish to solve (\ref{eq:EKG1}), meaning we need to ensure $\xi=0$ on solutions of (\ref{eq:EinsteinDeTurck}). We are interested in stationary, axisymmetric spacetimes with a $t-\phi$ reflection symmetry which, according to \cite{headrick2010new,figueras2011ricci}, give a second order system of elliptic PDEs. Furthermore, it has been shown that, when $\psi=0$, solutions with $\xi^a\neq0$ cannot exist \cite{figueras2016existence}. For the case at hand though, due to the presence of a scalar field, we have to verify \emph{a posteriori} that this is the case. Since the equations are elliptic, local existence theorems imply that a solution with $\xi\neq0$ cannot be arbitrarily close to one with vanishing $\xi$.

The most generic \emph{ansatz} for such a spacetime is
\begin{multline}
\mathrm{d}s^2=-F(x,z)x^2\mathrm{d}t^2+r_0^2\bigg\{\frac{4C(x,z)}{(1-x^2)^4}\mathrm{d}x^2\\
+\frac{A(x,z)(1-z^2)^2}{(1-x^2)^2}\left[\mathrm{d}\phi-(1-x^2)^2W(x,z)\frac{\mathrm{d}t}{r_0}\right]^2\\
+\frac{4D(x,z)}{(1-x^2)^2(2-z^2)}\left[\mathrm{d}z+B(x,z)\mathrm{d}x\right]^2\bigg\},
\label{eq:HBHansatz}
\end{multline}
where $r_0$ is the BH radius. Here, $x\in(0,1)$ plays the role of a radial coordinate with $x=0$ being the horizon and $x=1$ asymptotic spatial infinity. $z\in(-1,1)$ is an angular coordinate, with $z=-1$ being the south pole of the horizon and $z=1$ the north. There is also a $\mathbb{Z}_2$ reflection symmetry $z\to-z$, so we will take $z\in(0,1)$ and impose relection symmetry at $z=0$.

Appropriate boundary conditions have to be imposed at the edges of our domain and these can be obtained by expanding the equations of motion \eqref{eq:EKG2},\eqref{eq:EinsteinDeTurck} about the corresponding boundaries \cite{Supp}. Our choice of reference metric is based on obtaining the Kerr metric asymptotically, which amounts to $A=F=C=D=1$, $B=0$ and $W=\widehat{\Omega}(1-x^2)$, so that its angular velocity vanishes at infinity and is fixed at the horizon to $\Omega_{H}=\widehat{\Omega}/r_0$. Using the boundary conditions at the horizon, one can show that the temperature of the hairy solution is $T_{H}=1/(4\pi\,r_0)$. Since general relativity has no scale, we will measure all physical quantities in temperature units.

Finally, for the scalar field we take
\begin{align}
\psi(t,x,z,\phi)=e^{-i\tilde{m}\Omega_{H} t}e^{i\tilde{m}\phi}(1-z^2)^{\tilde{m}}\,\tilde{\psi}(x,z)\,,
\end{align}
where $(1-z^2)^{\tilde{m}}$ ensures regularity of the scalar field at the south and north poles \cite{Supp}. At asymptotic infinity we demand that $\tilde{\psi}=0$ and at the horizon we require regularity, which is enforced via $\left.\partial \tilde{\psi}/\partial x\right|_{x=0}=0$.

The moduli space of solutions is then generated by varying $\Omega_{H}$ and the integer $\tilde{m}$\footnote{In the actual code we treat $\Omega_{H}$ as an unknown number and provide an additional equation in the form of a very small normalisation condition $\tilde{\psi}(0,1)=\epsilon$ on the scalar field at the horizon ($x=0$, $z=1$). The moduli space of solutions is then generated by varying $\epsilon$ at fixed integer $\tilde{m}$.}. Our numerical findings for the background are consistent with those in \cite{herdeiro2014kerr}.
\\
\indent{\bf~Perturbing the HBHs --} In order to investigate the stability of the HBHs, we have to perturb \eqref{eq:EKG}. We consider small changes in both the metric and scalar field:
\begin{subequations}
\begin{align}
g_{ab}&= g_{ab}^{(0)}+h_{ab}\,,
\\
\psi&=\psi^{(0)}+\eta\,,
\end{align}
\label{eq:perturbs}%
\end{subequations}
where ${}^{(0)}$ represents background quantities. These give rise to the following equation for the perturbed scalar:
\begin{subequations}
\begin{align}
&\Box^{(0)}\eta-\mu^2\eta-\hat{L}^{(0)}\psi^{(0)}=0
\label{eq:KGperturb}
\\
&\hat{L}^{(0)}=\bar{h}^{ab}\nabla^{(0)}_{\vphantom{b} a}\nabla^{(0)}_b-\nabla^{(0)}_{\vphantom{d} a}\bar{h}^{ad}\nabla^{(0)}_d-\frac{1}{2}\mu^2\bar{h},
\end{align}
\end{subequations}
where we have also defined the trace-reversed metric perturbation $\bar{h}_{ab}\equiv h_{ab}-\frac{1}{2}h\,g^{(0)}_{ab}$ and $h\equiv g^{(0)}_{ab}h^{ab}$.

Next, we have to choose a way to fix the gauge freedom induced by the following transformations
\begin{subequations}
\begin{align}
&h_{ab}\rightarrow h_{ab}+\mathcal{L}_{\chi}g^{(0)}\,,
\label{eq:GaugeTransa}
\\
&\eta\rightarrow\eta+\mathcal{L}_\chi\psi^{(0)}\,,
\label{eq:GaugeTransb}
\end{align}
\label{eq:GaugeTrans}%
\end{subequations}
where $\chi$ is assumed to be the same order as $h_{ab}$ and $\eta$.

Even though, one would like to completely separate the scalar from the gravitational perturbations in the EKG equations \eqref{eq:EKG}, this does not seem possible in our case.

The most we can achieve is to choose a gauge in such a way as to decouple the perturbed Klein-Gordon equation from the metric perturbations $h_{ab}$, while still leaving the perturbed Einstein equation sourced by the scalar perturbation $\eta$. One way of doing this is by first setting:
\begin{equation}
\nabla^{(0)}_a\bar{h}^{ad}=P^d(\bar{h},\bar{h}_{ab})\,.
\label{eq:conditiongauge1}
\end{equation}
It is essential for our proof that $P_d$ can only depend on $h_{ab}$ but not its derivatives. In order to prove that such a gauge can be achieved, independently of our choice of $P$, we transform Eq.~(\ref{eq:conditiongauge1}) using Eqs.~(\ref{eq:GaugeTrans}):
\begin{equation}
\Box_{\vphantom{d}}^{(0)}\chi_d+R^{(0)}_{da}\chi^a+\nabla^{(0)}_{\vphantom{d} a}\bar{h}^a_{\phantom{a}d}-P_d-P^{(\chi)}_d=0\,,
\label{eq:GaugedCond}
\end{equation}
where $P^{(\chi)}_d$ is the gauge transformed version of $P_d$, which again depends only on $\bar{h}_{ab}$ but not on its first derivatives.

Moreover, \eqref{eq:GaugeTrans} tells us that $P^{(\chi)}_d$ also can only depend on first order derivatives of $\chi_d$, implying that the principal symbol of Eq.~(\ref{eq:GaugedCond}) is governed by $\Box^{(0)}$. We can then use Theorem $10.1.2$ of \cite{wald2010general} to show that $\chi$ can be chosen in such a way, as to have the above equation \eqref{eq:GaugedCond} uniquely satisfied for each component of $\chi$. This confirms that we can set $\nabla^{(0)}_a\bar{h}^{ad}=P^d$ for any choice of $P^d$ that can depend at most on $h_{ab}$, but not on its derivatives.

With this gauge choice, we would like to set $\hat{L}^{(0)}\psi=0$ in \eqref{eq:KGperturb}, which will decouple the perturbed KG equation from the metric perturbations $\bar{h}_{ab}$. This translates to being able to uniquely solve
\begin{equation}
\bar{h}^{ab}\nabla_a\nabla_b\psi^{(0)}-P^d\nabla^{(0)}_d\psi_{\vphantom{d}}^{(0)}-\frac{1}{2}\mu^2\bar{h}\psi^{(0)}=0.
\label{eq:conditiongauge2}
\end{equation}
This can then be easily achieved by choosing $P^d$, and noting again that, as desired, it turns out to be a function of $\bar{h}_{ab}$ only. The final equation to be solved is then
\begin{equation}
\Box^{(0)}\eta-\mu^2\eta=0\,.
\label{eq:perturbfinal}
\end{equation}

Finally, we come to the thorny issue of residual gauge transformations $\hat{\chi}$, \emph{i.e.} gauge transformations that leave the gauge condition (\ref{eq:conditiongauge1}) invariant. One can show that such residual gauge transformations necessarily satisfy
$$
\Box^{(0)}\hat{\chi}_d+R^{(0)}_{da}\hat{\chi}^a-P^{(\chi)}_d=0\,,
$$
We have to show that such gauge perturbations \emph{cannot} be used to set all solutions of (\ref{eq:perturbfinal}) to zero using Eq.~(\ref{eq:GaugeTransb}) with $\chi=\hat{\chi}$. We devise a test to distinguish pure gauge from physical modes based on the fact that the former necessarily produce a metric perturbation that diverges exponentially at large distances, thus becoming incompatible with the requirement of asymptotic flatness.

By performing a Frobenius analysis close to $x=1$ (asymptotic infinity) it can be shown that
\begin{equation}
\eta = e^{-\frac{\Gamma}{1-x}} (1-x)^{\kappa}\tilde{\eta}(t,x,\theta,\phi)
\end{equation}
where $\tilde{\eta}(t,x,\theta,\phi)$ is a polynomial in $(1-x)$ and $\Gamma\,,\kappa\in\mathbb{C}$. A similar analysis can be repeated for $\psi^{(0)}$ and gives
\begin{equation}
\psi^{(0)} = e^{-\frac{\tilde{\Gamma}}{1-x}}  (1-x)^{\tilde{\kappa}}\tilde{\psi}^{(0)}(t,x,\theta,\phi)\,,
\end{equation}
with $\tilde{\psi}^{(0)}(t,x,\theta,\phi)$ polynomial in $(1-x)$ and $\tilde{\Gamma}\,,\tilde{\kappa}\in\mathbb{R}$.

Assume momentarily $\tilde{\Gamma}>\mathrm{Re}(\Gamma)$. If $\eta$ is pure gauge, then from \eqref{eq:GaugeTransb} with $\chi=\hat{\chi}$, the residual gauge perturbation $\hat{\chi}$ has to blow up exponentially as $x\to1^-$. However, this generates a metric perturbation, via Eq.~(\ref{eq:GaugeTransa}), that necessarily diverges exponentially as $x\to1^-$, thus becoming inconsistent with the assumption of asymptotic flatness \cite{Arnowitt:1962hi,Ashtekar:1978zz,Ashtekar:1991vb}. Therefore, by comparing the behaviour of the numerically computed perturbations at asymptotic infinity to the decay of the scalar hair in our background solutions, we can say whether the mode has a chance of being pure gauge. Crucially, the above argument shows that modes with $\tilde{\Gamma}>\mathrm{Re}(\Gamma)$ are necessarily physical.

\indent{\bf~Results --} In order to solve Eq.~(\ref{eq:perturbfinal}) we take advantage of the fact that the background metric $g_{ab}^{(0)}$ is stationary and axisymmetric and as such that we can decompose the scalar field perturbation as $\eta = \hat{\eta}(x,z)e^{-i\,\omega\,t+i\,m\,\phi}$ and solve for $(\hat{\eta},\omega)$ given a value of $m$\footnote{Recall that $\tilde{m}$ denotes the azimuthal quantum number of the background solution, and $m$ the quantum number of the corresponding perturbations.}. To solve the resulting eigenvalue problem we will use Newton's method \cite{cardoso2013holographic}. For the numerical simulations we use spectral collocation methods on a Chebyshev grid and impose the appropriate boundary conditions given in the Supplemental Matetial \cite{Supp}.

We have computed the quasinormal mode spectrum of HBHs with $\tilde{m}=1$, close to the superradiance onset in Kerr (see Fig.1 in \cite{Degollado:2018ypf}), for perturbations with $m=1$ and $m=2$. The former turns out to be pure gauge ($\tilde{\Gamma}=\mathrm{Re}(\Gamma)$), corresponding to shifts in the phase space of HBHs - altering the mass and angular momentum of the scalar cloud around the BH. The modes with $m=2$, however, are physical and \emph{always} unstable in the regions where the $\tilde{m}=1$ HBHs exist.

\begin{figure}
\centering
	\includegraphics[width=.47\textwidth]{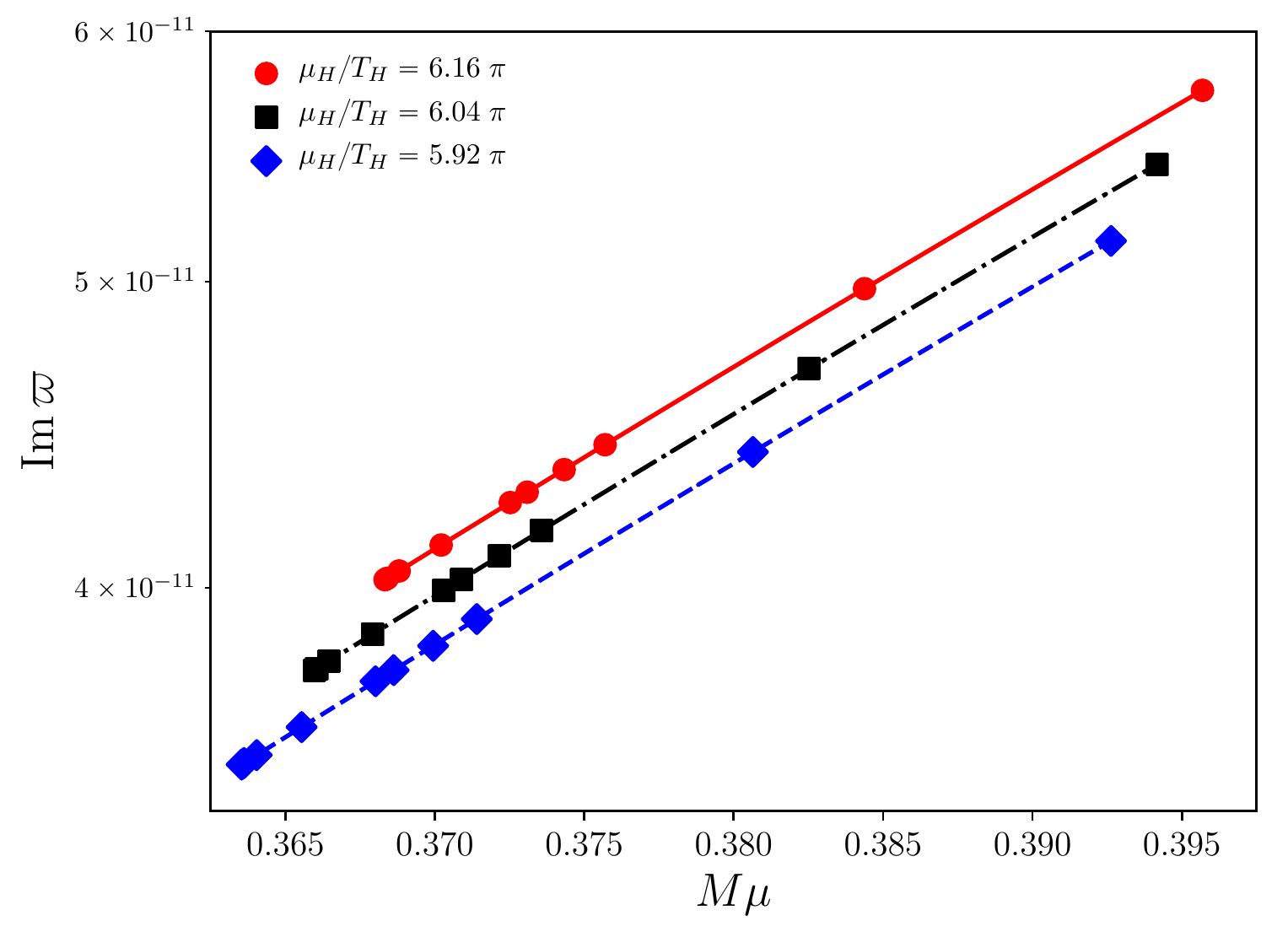}
	\caption{The imaginary part of $\varpi$ around HBHs, computed with $m=2$, as a function of $\mu M$ - each curve contains a family of HBHs with a fixed value of $\mu_H/T_H$.}
	\label{fig:imaginary}
\end{figure}

\begin{figure}
	\includegraphics[width=.47\textwidth]{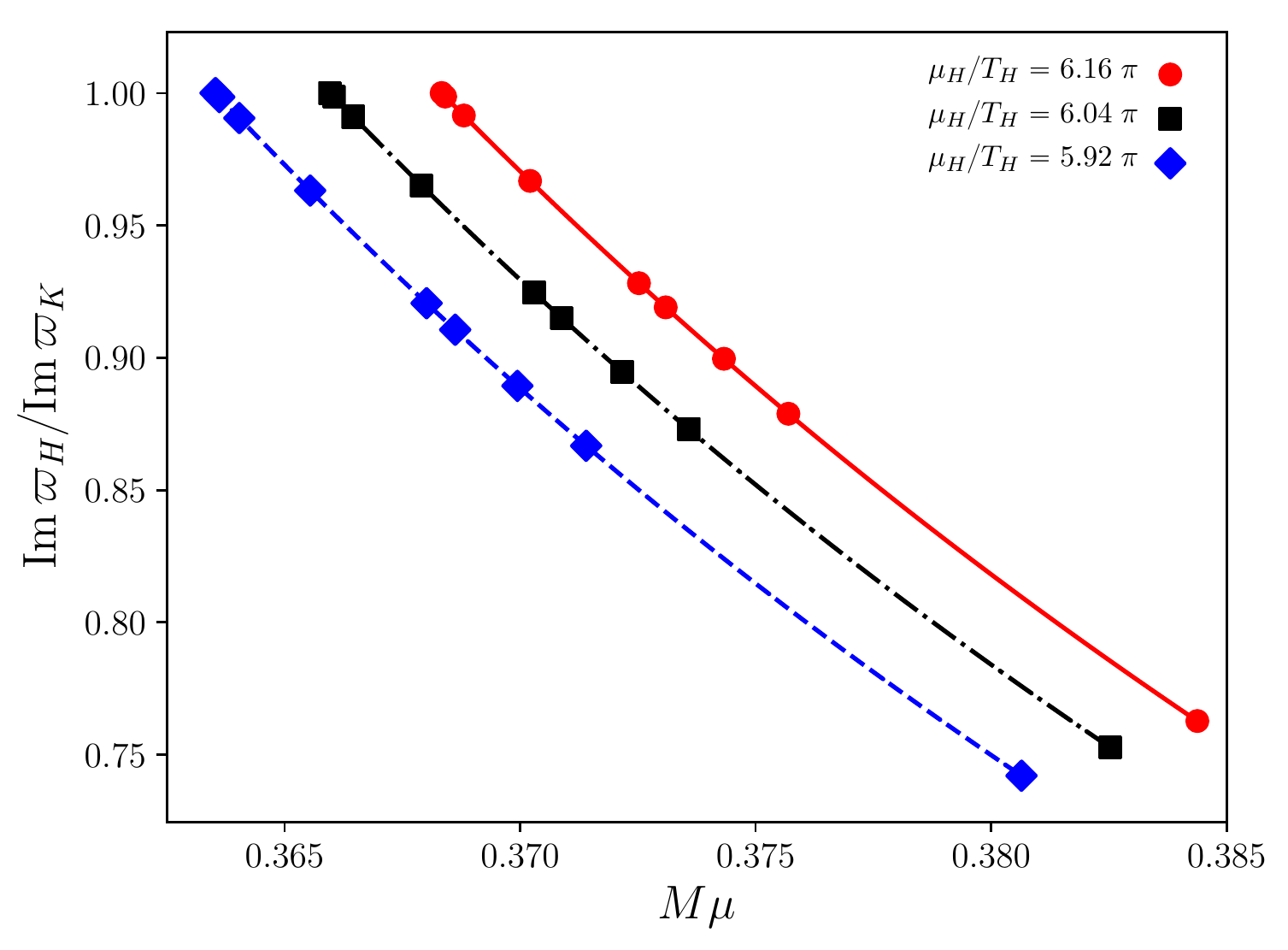}
	\caption{The ratio $\varpi_H/\varpi_K$, as a function of $\mu M$ - each curve contains a family of HBHs with a fixed value of $\mu_H/T_H$ and Kerr BHs with the same $J \mu^2$ and $M\mu$ as the HBHs.}
	\label{fig:imaginaryMu}
\end{figure}

This is summarised in Fig.~\ref{fig:imaginary}, where we plot $\varpi \equiv \omega/\mu$ as a function of $M\,\mu$. Each curve represents a different constant value of $\mu/T$ for the background HBHs. In order to compare the growth rate of the instability with that of a Kerr BH with the \emph{same} dimensionless angular momentum $J \mu^2$ and dimensionless mass $M\mu$, we plot in Fig.~\ref{fig:imaginaryMu} the ratio $\varpi_H/\varpi_K$, with $\varpi_H$ being computed using the HBHs and $\varpi_K$ with a Kerr BH with the same $J \mu^2$ and $M\mu$. The fact this ratio is always below unity, indicates that the HBHs are less unstable at fixed mass, angular momentum and scalar mass, as argued in \cite{Degollado:2018ypf}.

In the Supplemental Material \cite{Supp}, the imaginary parts of the above presented QNM spectra are plotted separately against $J/M^2$, demonstrating their positivity for the range of parameters considered and clearly showing that the rightmost point on each curve in Fig.~\ref{fig:imaginaryMu} lies in the region of superextremality $J/M^2>1$, where Kerr BHs do not exist. We anticipate similar results for higher $m$ modes. The real parts of the QNM spectra for perturbations around HBHs are shown in the Supplemental Material \cite{Supp}.
\\
\indent{\bf~Conclusions --} We have perturbed the HBHs of \cite{herdeiro2014kerr} and shown that they are unstable to linear mode perturbations. For the range of parameters that we have analysed, these BHs are uniquely identified by an integer $\tilde{m}$ and by the ratios $\mu/T_H$ and $J/M^2$ (or $\Omega_{H}/T_{H}$). All unstable modes we found have $m>\tilde{m}$. Furthermore, for small amplitudes of the scalar hair around the BH the growth rate of the instability is comparable to that of a massive scalar field around a member of the Kerr family, whereas for large amplitudes the HBHs are a few times less unstable than their nonhairy counterparts.

By comparing the growth rates of the fastest and slowest growing  $m=1$ modes around Kerr, with their equivalent $m=2$ modes around a HBH at the same $J/M^2$ and fixed gravitational coupling $M\mu$, we can assess the astrophysical significance of the portion of the moduli space of HBHs of \cite{herdeiro2014kerr} we studied. In this comparison we are neglecting the energy radiated during the formation of the HBH, which is a reasonable approximation \cite{Brito:2014wla,Brito:2017zvb,East:2017ovw,East:2017mrj}. We take a HBH with $J/M^2=0.983$ and $M\mu = 0.3844$, where the fastest decay is observed and another one with $J/M^2=0.946$ and $M\mu = 0.3635$, where the instability is the weakest. Our data imply that the former undergoes an instability, due to the $m=2$ mode, evolving on timescales between $\tau\sim 4.8\times 10^6$ s and $\tau\sim 1.7\times 10^7$ s for the smallest and largest final mass BHs as detected by LIGO-VIRGO and $\tau\sim 2.7\times 10^{15}$ s for supermassive BHs ($10^{10}\,M_\odot$). For the corresponding Kerr BH, the $m=1$ superradiant mode extracts energy efficiently \cite{Brito:2014wla,Brito:2017zvb,East:2017ovw,East:2017mrj}, exhibiting e-folding times between $\tau\sim 795$ s and $\tau\sim 2740$ s, for the same intermediate masses as above, and $\tau\sim 4.4\times 10^{11}$ s in the case of supermassive BHs. The second HBH is subject to instabilities with lifetimes of the order of $\tau\sim 6.9\times10^6$ s, $\tau\times 2.4\times10^7$ s and $\tau\times 3.8\times10^{15}$ s for the three cases of BH masses. The complementary unstable Kerr solution experiences similar rates - $\tau\sim 1.1\times 10^7$ s, $\tau\sim 3.7\times10^7$ s and $\tau\sim 6.0\times10^{15}$ s accordingly. This implies that in the explored region of parameter space HBHs may suffer from superradiance on the same scale as their nonhairy counterparts, but can also be distinctly more robust to its effects. Nevertheless, the timescales involved in both processes are at least two orders of magnitude smaller than the age of the universe, for the HBH solutions we analysed.

We should note that some regions of the moduli space of the HBHs are not excluded by our analysis \cite{Degollado:2018ypf}. In particular, it has been predicted in \cite{Kleihaus:2007vk,Herdeiro:2014jaa}, that starting from the onset of superradiance in Kerr, and continuing along a line of constant $M\mu$, the instability growth rate for HBHs decreases towards zero as the corresponding ergoregionless-boson-star is approached. Thus, it is conceivable that the region not excluded by our analysis is actually larger than the excluded region.

\indent{\bf~Acknowledgments --} B.G. would like to thank C.~Ganguly and T.~Bjorkmo for helpful discussions. The authors would also like to thank C.~Herdeiro for an insightful email exchange after the publication of \cite{Degollado:2018ypf}. We also thank O.~J.~C.~Dias for reading an earlier version of this manuscript and H.~S.~Reall for insightful discussions. B.G. was supported by an STFC studentship. J.E.S. was supported in part by STFC Grants No. PHY-1504541 and ST/P000681/1.

\bibliography{Paperbib}
\bibliographystyle{unsrt}
\newpage
\section{Supplemental Material}

\indent{\bf~Perturbed Einstein equation --} We can also perturb the Einstein equation
\begin{equation}
\frac{R_{ab}}{8\pi}=2\nabla_{(a}\psi^*\nabla_{b)}\psi+g_{ab}\mu^2\psi^*\psi\,,
\label{eq:EKG1Supp}
\end{equation}
with 
\begin{subequations}
	\begin{align}
	g_{ab}&= g_{ab}^{(0)}+h_{ab}\,,
	\\
	\psi&=\psi^{(0)}+\eta\,,
	\end{align}
	\label{eq:perturbsSupp}%
\end{subequations}
trace-reverse the resulting equations and utilise the Ricci identitiy
\begin{align}
2\nabla^{(0)}_{\left[c\right.}\nabla^{(0)}_{\left.d\right]}V_{ab}=R^{(0)}_{becd}\tensor{V}{_a^e}-R^{(0)}_{eacd}\tensor{V}{^e_b}
\end{align}
in order to obtain
\begin{multline}
\tensor*{R}{^{(0)}_{c(a}}\tensor{\bar{h}}{_{b)}^{\vphantom{)} c}}+\tensor*{R}{^{(0)}_{dbac}}\tensor{\bar{h}}{^{dc}}+\nabla^{(0)}_{(a\rvert}\nabla^{(0)}_{\vphantom{(} c}\bar{h}_{\lvert b)}^{\vphantom{)} c}-\nabla^{(0)}_{\vphantom{b} a}\nabla^{(0)}_b\bar{h}-\ \\
\frac{1}{2}\Box^{(0)}\bar{h}_{ab}+\frac{1}{4}g^{(0)}_{ab}\Box^{(0)}_{\vphantom{b}}\bar{h}=8\pi\bigg[2\nabla^{(0)}_{(a}\eta^*\nabla^{(0)}_{b)}\psi^{(0)}_{\vphantom{)}}+\\
2\nabla^{(0)}_{(a}\big(\psi^{(0)}\big)^*\nabla^{(0)}_{b)}\eta+g_{ab}\mu^2\left(\eta^*\psi^{(0)}+\big(\psi^{(0)}\big)^*\eta\right)\bigg].
\end{multline}
Unfortunately, due to the presence of matter, the equations cannot be simplified any further, unlike in the vacuum case, where the Ricci tensor vanishes and tracing the above equation leads to $\Box^{(0)}\bar{h}=0$. Fortunately, due to the decoupling of the KG equation for $\eta$ from the metric perturbations, we do not need to know the form of $h_{ab}$ in order to assess the linear stability of the HBHs.

\indent{\bf~Boundary conditions --} First, we present the required boundary conditions for the construction of HBHs using
\begin{multline}
R_{ab}-\nabla_{\left(a\right.}\xi_{\left.b\right)}=8\pi \left[2\nabla_{(a}\psi^*\nabla_{b)}\psi+g_{ab}\mu^2\psi^*\psi\right],
\label{eq:EinsteinDeTurckSupp}
\end{multline}
with ansatz
\begin{multline}
\mathrm{d}s^2=-F(x,z)x^2\mathrm{d}t^2+r_0^2\bigg\{\frac{4C(x,z)}{(1-x^2)^4}\mathrm{d}x^2\\
+\frac{A(x,z)(1-z^2)^2}{(1-x^2)^2}\left[\mathrm{d}\phi-(1-x^2)^2W(x,z)\frac{\mathrm{d}t}{r_0}\right]^2\\
+\frac{4D(x,z)}{(1-x^2)^2(2-z^2)}\left[\mathrm{d}z+B(x,z)\mathrm{d}x\right]^2\bigg\},
\label{eq:HBHansatzSupp}
\end{multline}
and a reference metric with line element as above with particular values for the unknown functions given as $A=F=C=D=1$, $B=0$ and $W=\widehat{\Omega}(1-x^2)$, with details on this choice presented in the main text.\\
At the horizon ($x=0$), requiring regularity, we set
\begin{align}
&\partial_xA=\partial_xD=\partial_xC=\partial_x\psi=0,\notag\\
&F=C,\;W=\Omega_{HBH},\;B=0\,.
\end{align}
At asymptotic infinity ($x=1$), where the metric has to approach Minkowski spacetime, we have 
\begin{align}
&A=C=D=F=1\notag\\
&W=B=\psi=0\,.
\end{align}
At the north pole ($z=1$), also demanding regularity, we get
\begin{align}
&\partial_zF=\partial_zD=\partial_zC=\partial_zW=\partial_z\psi=0,\notag\\
&A=D,\;B=0\,.
\end{align}
And finally at the axis of the polar angle reflection symmetry ($z=0$), insisting on smoothness, we require
\begin{align}
&\partial_zA=\partial_zF=\partial_zD=\partial_zC=\partial_zW=\partial_z\psi=0,\notag\\
&B=0\,.
\end{align}

\indent{\bf~Boundary conditions for perturbations --} 
In this section, and without loss of generality, we set $r_0=1$. We now look at the Klein-Gordon equation on a fixed Kerr or hairy background. Apart from imposing boundary conditions it is also necessary to factor out the divergent behaviour of the scalar field at the boundaries of the coordinate grid. The required factors can be inferred by performing a Frobenius analysis about the locations of interest. Near asymptotic infinity ($x=1$) we have a wave-like equation in the radial direction, with the constant term coming at a different order in the expansion, suggesting a factor of the form $e^{\frac{\alpha}{1-x^2}}$. Furthermore, the $z$-derivatives show up earlier in the series than the radial ones, requiring a further $(1-x^2)^\beta$ to be taken out. $\alpha$ and $\beta$ are determined by the series expanded equations and the requirement for a finite energy solution. The boundary condition itself turns out to be of Robin type.
\begin{align}
&\hat{\eta}(x,z)=e^{\frac{\alpha}{1-x^2}}(1-x^2)^\beta f(x,z)\,,\notag\\
&\alpha=-\sqrt{\mu^2-\omega^2}\,,\notag\\
&\beta=1+\frac{2\mu^2-4\omega^2+(\mu^2-\omega^2)\partial_xC(1,z)+\omega^2\partial_xF(1,z)}{4\sqrt{\mu^2-\omega^2}}\,,\notag\\
&\left.\vphantom{\hat{U}}(1-z)\partial_xf(x,z)\right|_{x=1}=\left.\hat{U}f(x,z)\right|_{x=1}\,,\notag\\
&\hat{U}(x,z)=\left[G_1(z)\partial_{zz}+G_2(z)\partial_z+G_3(x,z)\right]\,,
\end{align}
with
\begin{gather}
G_1(z)=-\frac{(1-z)(2-z^2)}{4\sqrt{\mu^2-\omega^2}},\,\notag\\ G_2(z)=-\frac{z(3z^2-4m(2-z^2)-5)}{4(1+z)\sqrt{\mu^2-\omega^2}}\,,
\end{gather}
\begin{gather}
G_3(x,z)=-\frac{(1-z)}{4}\left(\partial_xA(x,z)+\partial_xD(x,z)\right)\notag\\
+\frac{(1-z)(3\mu^2-5\omega^2)}{8\sqrt{\mu^2-\omega^2}}\;\partial_xC(x,z)\notag\\
-\frac{1}{16}(1-z)\sqrt{\mu^2-\omega^2}\;\left(\partial_xC(x,z)\right)^2\notag\\
+\frac{(1-z)\sqrt{\mu^2-\omega^2}}{4}\;\partial_{xx}C(x,z)\notag\\
+\frac{(1-z)\omega^2}{8\sqrt{\mu^2-\omega^2}}\;\partial_xC(x,z)\;\partial_xF(x,z)\notag\\
-\frac{(1-z)(5\omega^2-\mu^2(7\omega^2-2\sqrt{\mu^2-\omega^2}))}{8\left(\mu^2-\omega^2\right)^{3/2}}\;\partial_xF(x,z)\notag\\
-\frac{(1-z)\omega^2(4\mu^2-3\omega^2)}{16\left(\mu^2-\omega^2\right)^{3/2}}\;\left(\partial_xF(x,z)\right)^2\notag\\
+\frac{(1-z)\omega^2}{4\sqrt{\mu^2-\omega^2}}\;\partial_{xx}F(x,z)\notag\\
+\frac{1-z}{4\left(\mu^2-\omega^2\right)^{3/2}}\left[\vphantom{\left(\sqrt{\mu^2}\right)}3\mu^4+8\omega^4+4m(1+m)(\mu^2-\omega^2)\right.\notag\\
\left.-2\mu^2\left(6\omega^2-\sqrt{\mu^2-\omega^2}\right)\right]
\end{gather}

Close to the horizon ($x=0$) the power series indicate the presence of a regular singularity, forcing us to pull out $x^\gamma$ in front, whereby the constant is determined by the expanded equations and the restriction to ingoing waves only at the horizon. We impose Neumann boundary conditions
\begin{align}
&\hat{\eta}(x,z)=x^\gamma f(x,z)\,,\notag\\
&\gamma=-2i(\omega-m\Omega_H)\,,\notag\\
&\left.\partial_xf(x,z)\right|_{x=0}=0\,.
\end{align}

At the north pole of the squashed sphere ($z=1$), the series expansion again signals for a regular singularity, necessitating a prefactor of $(1-z^2)^\delta$, with $\delta$ determined by the equations. The boudnary conditions are Neumann again
\begin{align}
&\hat{\eta}(x,z)=(1-z^2)^\delta f(x,z)\,,\notag\\
&\delta=m\,,\notag\\
&\left.\partial_zf(x,z)\right|_{z=1}=0\,.
\end{align}

In the neighbourhood of the symmetry axis $z=0$ we do not expect any singular behaviour, as this is not a true boundary\footnote{Had we sticked with the original range of the coordinates, the $z=-1$ boundary would've required the same treatment as $z=1$.}, and the series expanded KG equation confirms that. The reflection symmetry in the polar coordinate separates the physical states of the scalar field into two equivalent subsets and the choice of Neumann or Dirichlet boundary conditions, which we are free to make, selects one of the two. We choose the former
\begin{align}
\left.\partial_z\hat{\eta}(x,z)\right|_{z=0}=0\,.
\end{align}

\indent{\bf~Behaviour around poles --} Near the poles of the squashed sphere, $z=\pm1$, spherical symmetry is almost perfectly recovered, hence the angular part of the metric to lowest order in $(1-z)$ can be written as
\begin{align}
ds^2_{z=\pm1}&\approx\frac{A(x,z)}{(1-x^2)^2}\left(dz^2+(1\pm z)^2d\phi^2\right)\notag\\
&=\frac{A(x,\tilde{z})}{(1-x^2)^2}\left(d\tilde{z}^2+\tilde{z}^2d\phi^2\right)\,,
\end{align}
where in the second line we have applied a shift $1\pm z\rightarrow\tilde{z}$. In this way we see that the metric takes the familiar form of 2D flat space in polar coordinates. The latter are, however, not regular at the origin (which after the shift in $z$ corresponds to the pole of the sphere), thus forcing us to change to Cartesian coordinates, so as to investigate the behaviour of the scalar field there. This is easily achieved by the following transformation
\begin{equation}
(\tilde{x},\tilde{y})=(\tilde{z}\,\cos\phi\\[\jot]\,,\tilde{z}\,\sin\phi\\[\jot])\,,
\label{eq:polarToFlat}
\end{equation}
which takes the $d\tilde{z}^2+\tilde{z}^2d\phi^2$ part of the metric to $d\tilde{x}^2+d\tilde{y}^2$. The scalar field, has the general form
\begin{align}
\psi(t,x,\tilde{z},\phi)=e^{-i\,\omega\,t}e^{i\,\tilde{m}\,\phi}f(x,\tilde{z})\,,
\end{align}
whose $e^{i\,\tilde{m}\,\phi}$ part can be rationalised under the change of variables \eqref{eq:polarToFlat}, depending on the value of $\tilde{m}$. Computing the first two cases $-$ $\tilde{m}=1$ and $\tilde{m}=2$ $-$ illustrates a simple general relation, which can be straightforwardly verified with the help of trigonometric identities - namely
\begin{align}
e^{i\,\tilde{m}\,\phi}=\left(\frac{\tilde{x}+i\tilde{y}}{\sqrt{\tilde{x}^2+\tilde{y}^2}}\right)^{\tilde{m}}\,.\label{eq:expPole}
\end{align}
Therefore, regularity at the poles fixes the polar angular dependence of the scalar field $\psi$ at least as $\left(1\pm z\right)^{\tilde{m}}$, in order to compensate for the denominator in \eqref{eq:expPole}.

\begin{figure}
	\includegraphics[width=.47\textwidth]{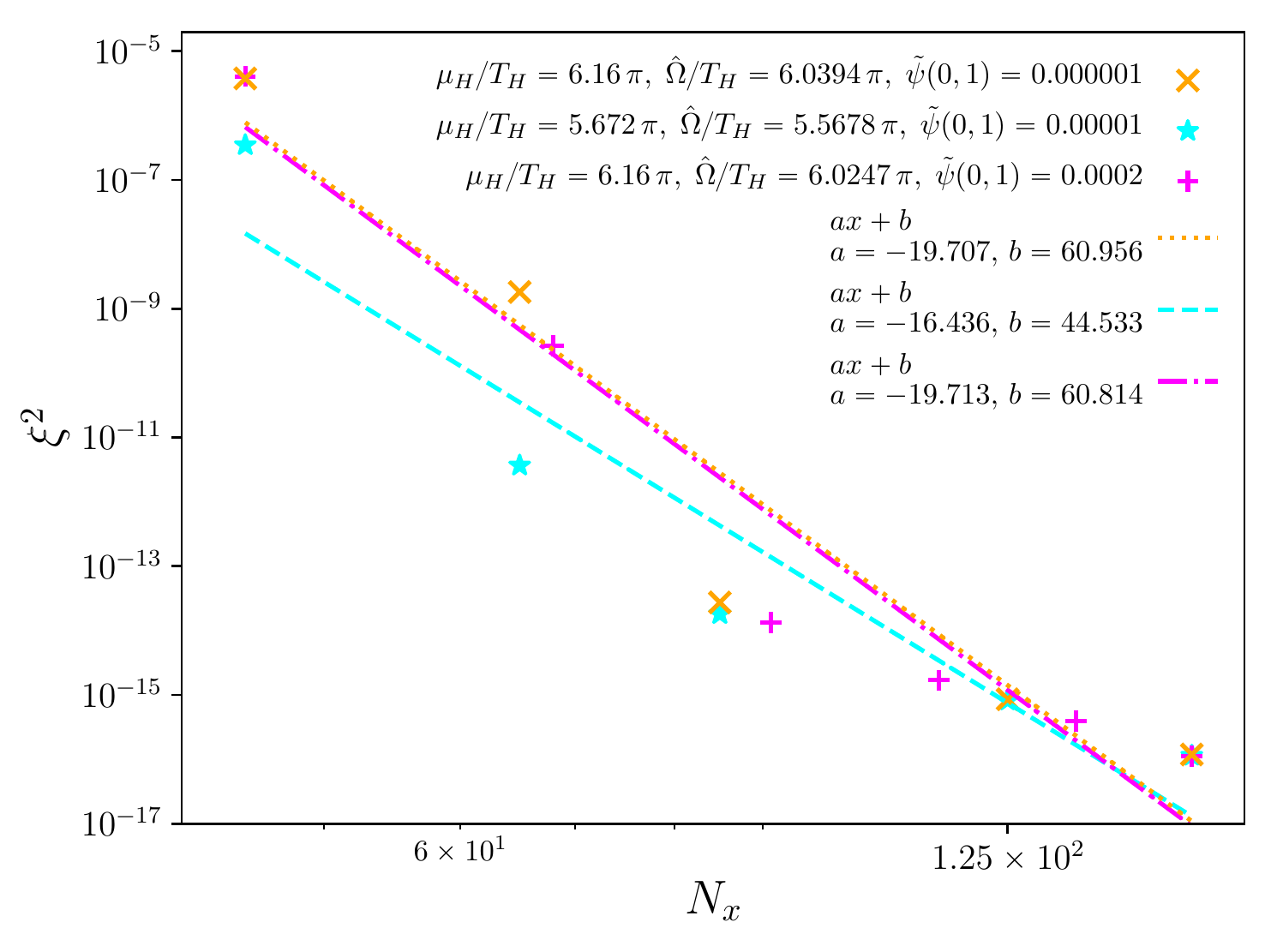}
	\caption{The maximum value of $\xi^2$ for the HBH solutions of (\ref{eq:EinsteinDeTurckSupp}) as a function of radial grid size $N_x$ on a Log-Log plot}
	\label{fig:DeTurckVec}
\end{figure}

\indent{\bf~Numerical tests --} In order to verify that solving (\ref{eq:EinsteinDeTurckSupp}) numerically gives us HBHs, we compute de norm of the DeTurck vector $\xi^2$ for each of the solutions. Here we will exhibit its convergence properties for three of the solutions, which we think should pose the biggest numerical challenges as they are the ones that stretch furthest into the corners of the HBH phase space that we have explored. This includes a solution with the highest value of the scalar field amplitude at the horizon which we have constructed, as well as the fastest and slowest spinning HBHs that we have managed to obtain. The results are shown in Fig.~\ref{fig:DeTurckVec} where we present the maximum value of $\xi^2$ for different radial grid sizes on a Log-Log plot. In the angular direction we are fixed at $N_z=35$, as this was found to be sufficient.
Only the lowest angular velocity solution (the first to be found) has been obtained at $N_x=125$. The maximum resolution in the radial direction is $N_x=160$. The closer to extremality, the harder it is to resolve the ${\rm AdS}_2$-like throat appearing near the horizon, thus the worse resolution for smaller grids.

To check that the superradiant modes that we compute can be trusted, we plot the ratio of the imaginary frequencies of solutions obtained at successive radial resolutions (the angular resolution is fixed at $N_z=35$)
\begin{align}
\left|1-\frac{{\rm Im}\,\omega_{K,N_x}}{{\rm Im}\,\omega_{K,N_x+\Delta}}\right|\,,
\end{align}
where $\Delta$ is the increase in the grid size - Fig.~\ref{fig:ScalarConvKerr}.
\begin{figure}
	\includegraphics[width=.47\textwidth]{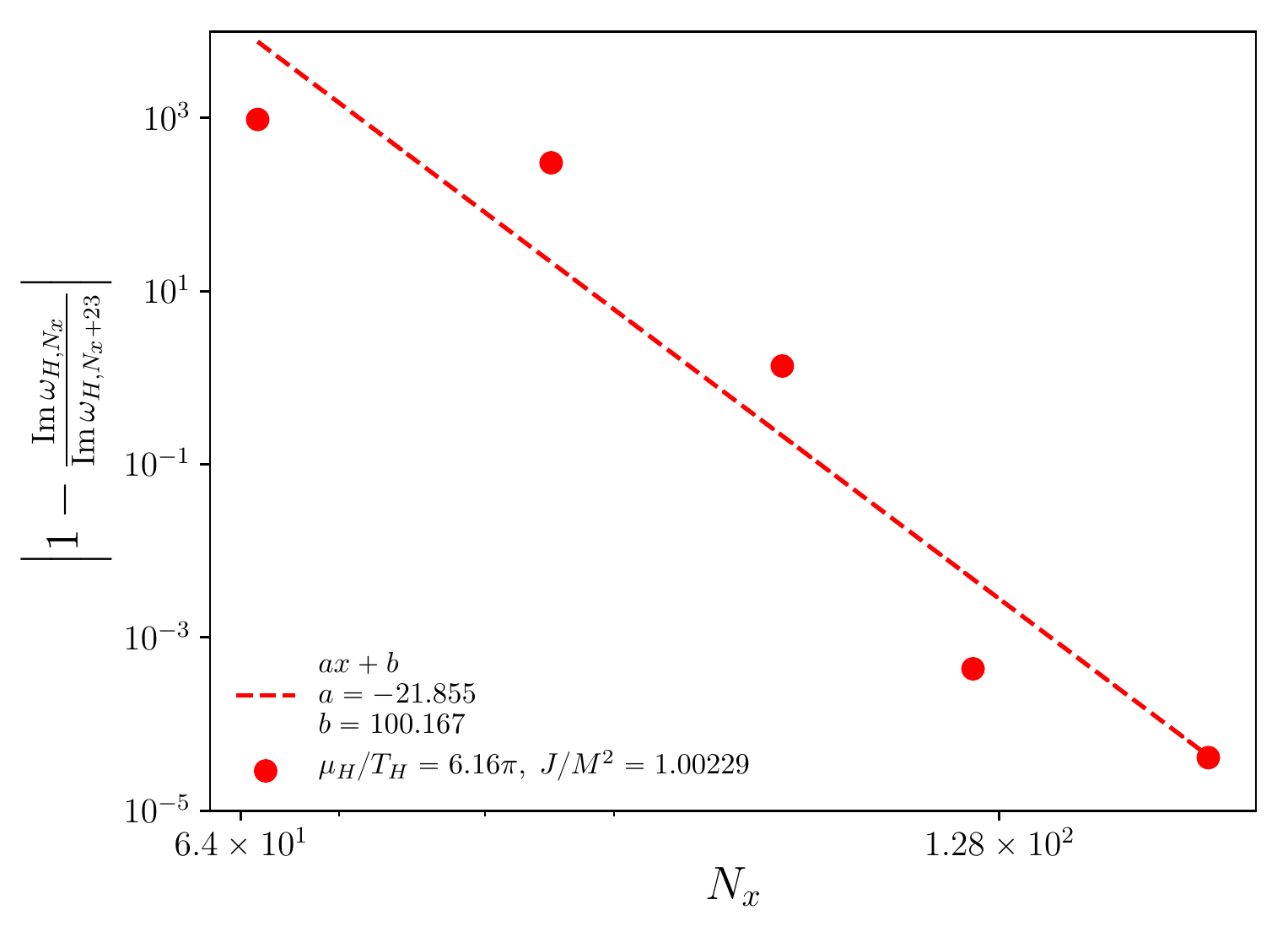}
	\caption{Numerical convergence of the imaginary part of the superradiant frequency of the scalar field with $m=2$ in a fixed HBH background for the highest scalar field amplitude at the horizon that we have considered in our studies.}
	\label{fig:ScalarConvHBH}
\end{figure}
We show results for a HBH background, as well as for the three fastest spininng Kerr BHs, due to their proximity to extremality, making them numerically challenging. Fig.~\ref{fig:ScalarConvHBH} is in the background of the HBH with the highest scalar field amplitude at the horizon from the solutions that we have found\footnote{The hardest one to work with from the HBHs.}, whereas Fig.~\ref{fig:ScalarConvKerr} represents the results in the three Kerr backgrounds discussed above. The decays are not exponential because of the non-analytic behaviour of the scalar field near asymptotic infinity ($x=1$) and the horizon ($x=0$). All results presented in the main section have been obtained at the highset resolution available for the respective spacetime - from $160\times 35$ to $260\times 35$, where we can safely trust the first four digits of the results.
\begin{minipage}[c]{0.5\textwidth}
	\centering
	\includegraphics[width=\textwidth]{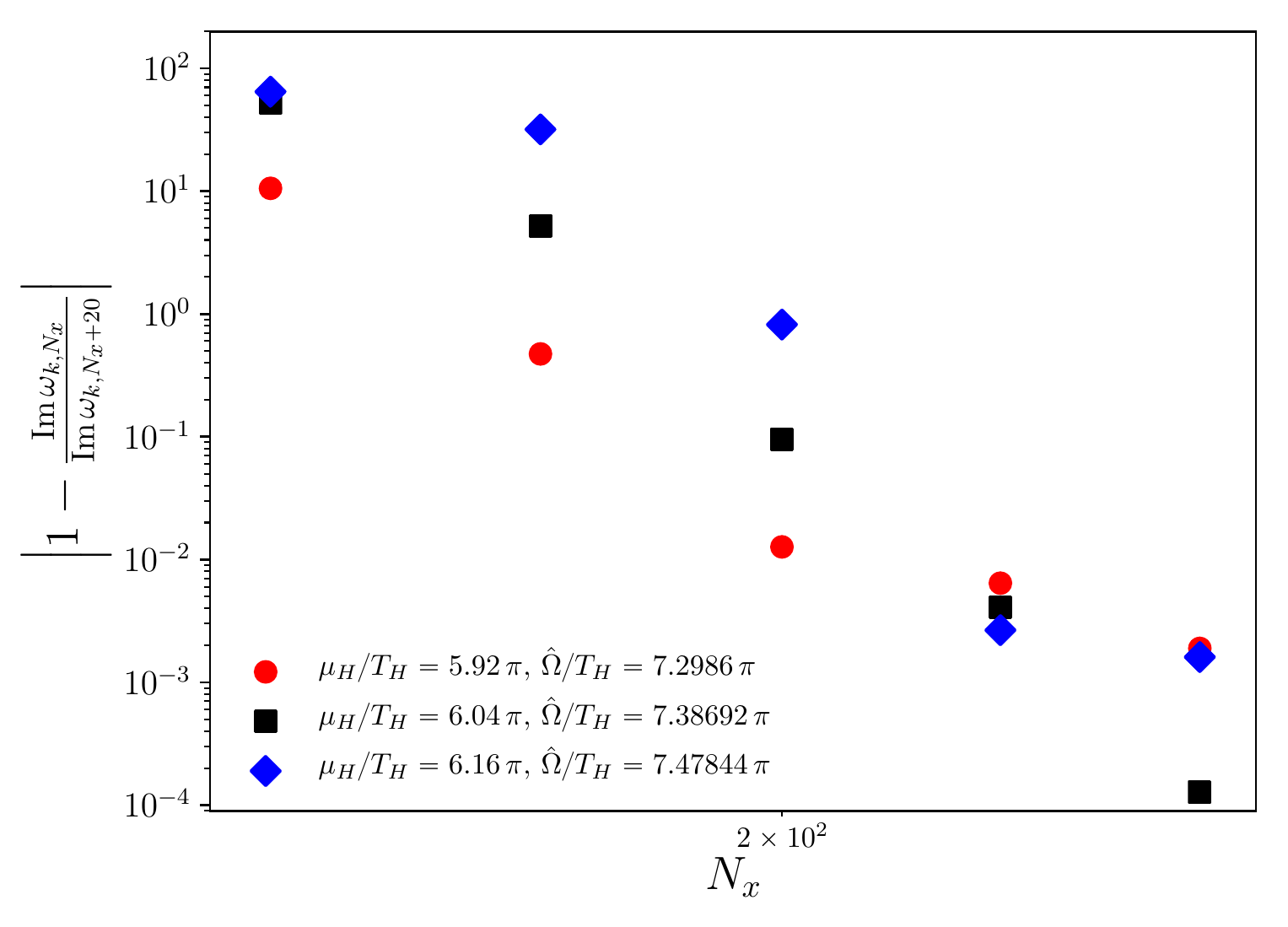}
	\captionof{figure}{Numerical convergence of the imaginary part of the superradiant frequency of the scalar field with $m=2$ in a fixed Kerr background for the three fastest spinning BHs that we have considered in our studies.}
	\label{fig:ScalarConvKerr}
\end{minipage}
\newpage
\indent{\bf~Real and imaginary parts of the QNM spectra of massive scalar field perturbations around hairy and Kerr black holes}
\begin{minipage}[c]{0.5\textwidth}
	\centering
	\includegraphics[width=\textwidth]{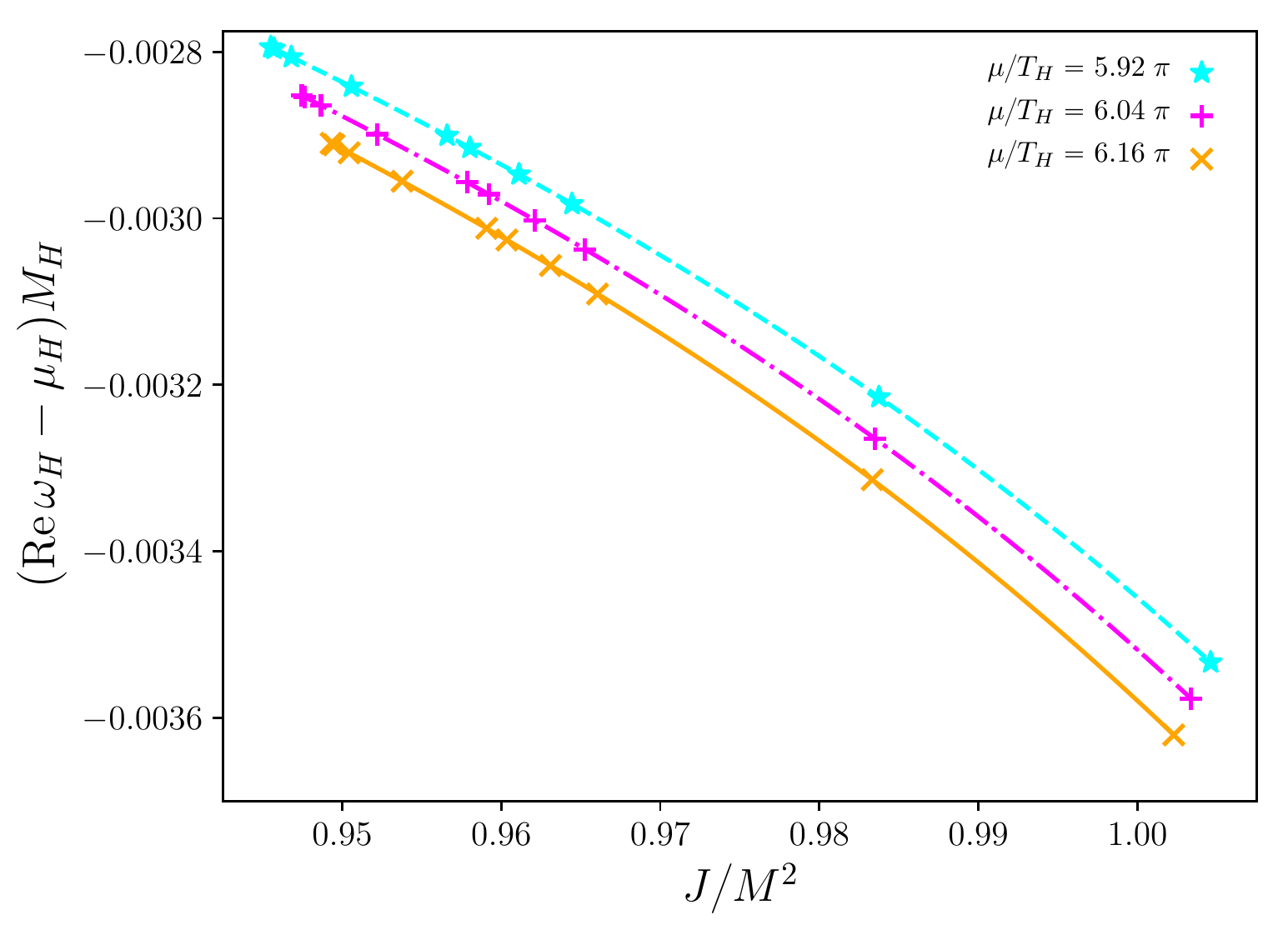}
	\captionof{figure}{Real part of the QNM spectra of massive scalar field perturbations, $m=2$, around HBHs ($\omega_H$) subtracted by the scalar mass ($\mu$) for several values of $\mu_H/T_H$ and $J/M^2$. $M_H$ - mass of the hairy background.}
	\label{fig:real}
\end{minipage}
\begin{minipage}[c]{0.5\textwidth}
	\centering
	\includegraphics[width=\textwidth]{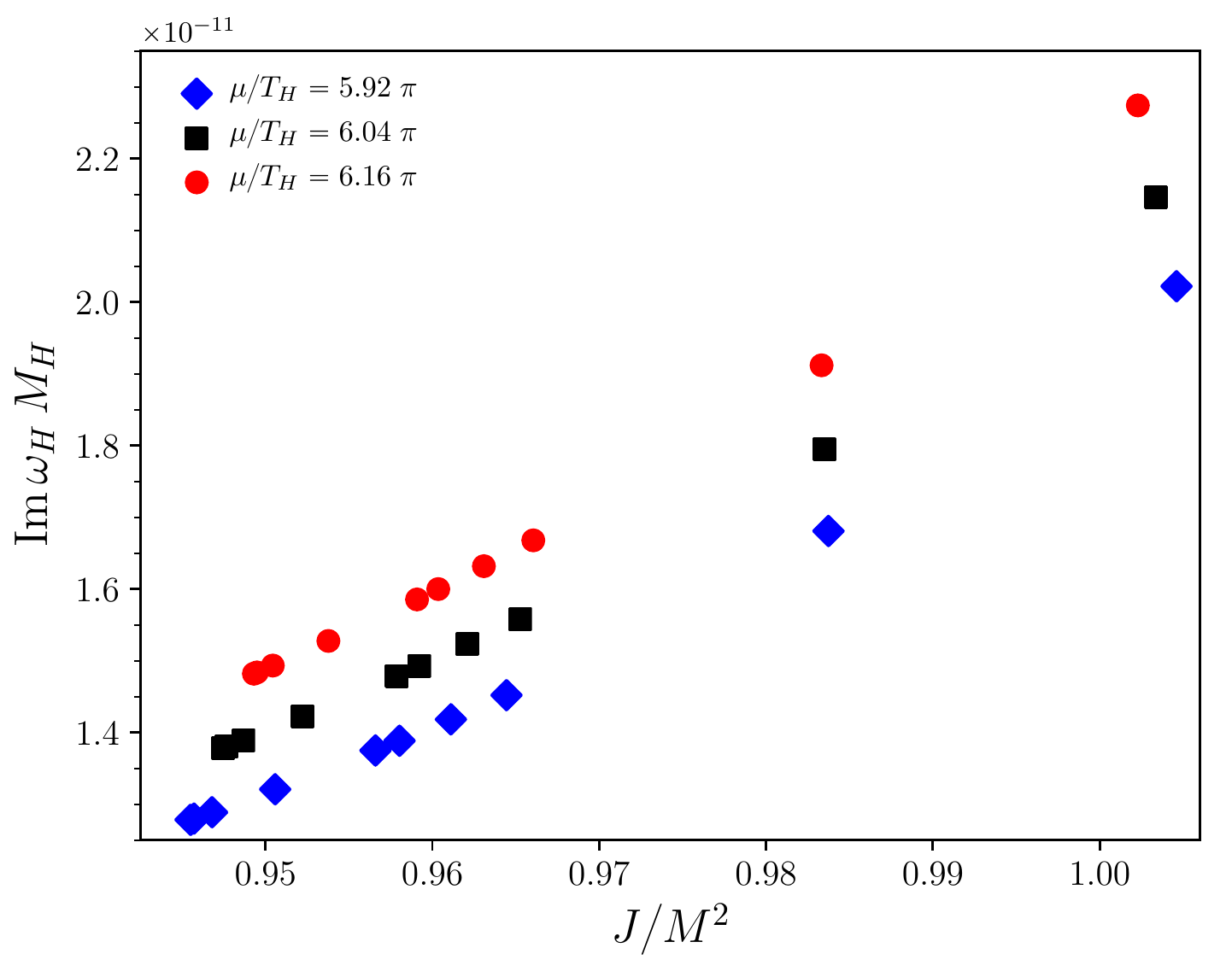}
	\captionof{figure}{The imaginary parts of the QNM spectra of massive scalar field perturbations, $m=2$, around hairy ($\omega_H$) BHs for several values of $\mu_H/T_H$ and $J/M^2$. $M_H$ is the mass of the respective HBH background. The rightmost point of each constant $\mu_H/T_H$ curve is in the superextremal region $J/M^2>1$, where Kerr black holes do not exist.}
	\label{fig:imaginaryOnlyHairy}
\end{minipage}
\begin{minipage}[c]{0.5\textwidth}
	\centering
	\includegraphics[width=\textwidth]{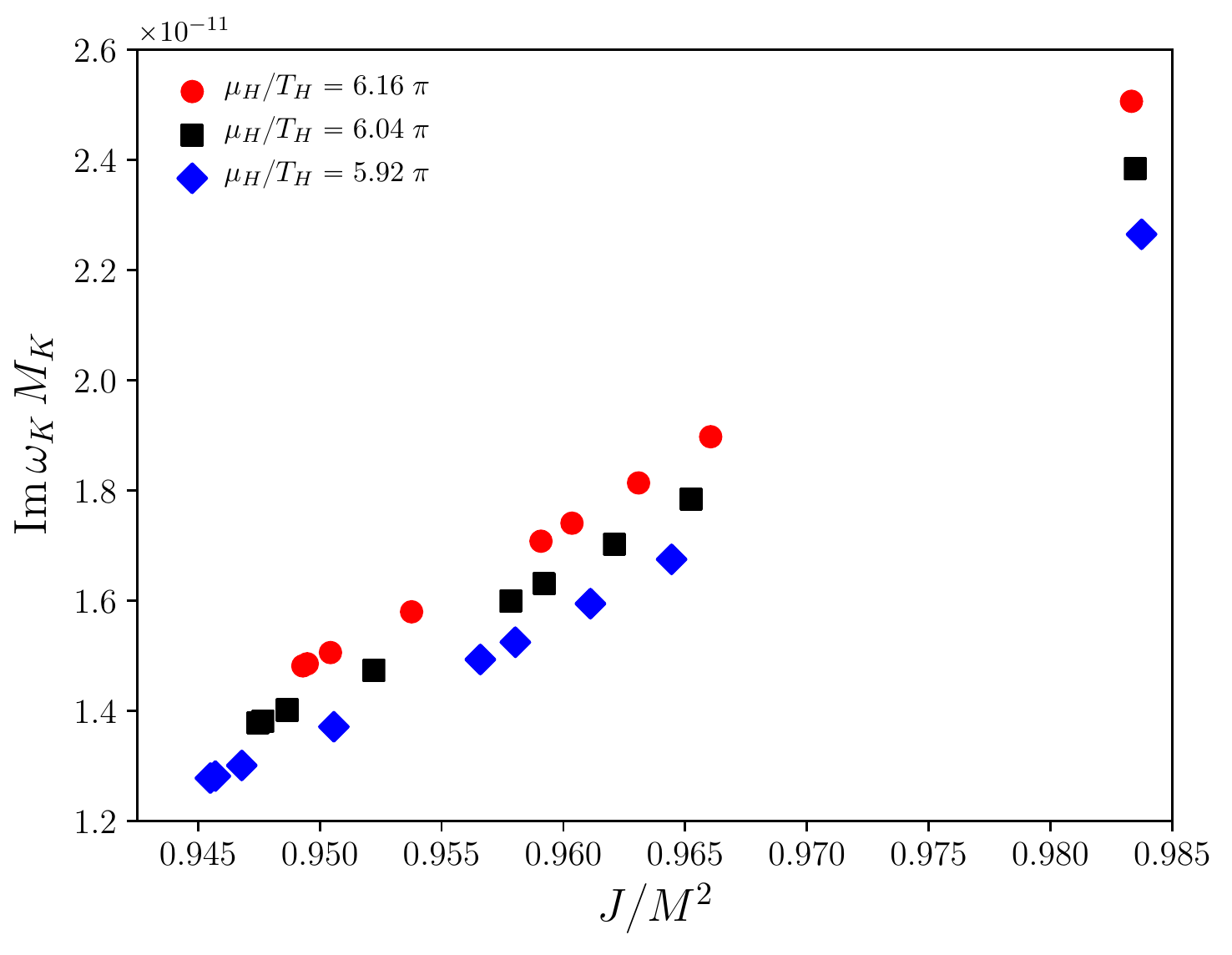}
	\captionof{figure}{The imaginary parts of the QNM spectra of massive scalar field perturbations, $m=2$, around Kerr ($\omega_K$) BHs as a function of $J/M^2$. Each background solution has the same $J\mu^2$ and $M\mu$ as the matching HBH point on Fig.~\ref{fig:imaginaryOnlyHairy}. $M_K$ is the mass of the respective Kerr background.}
	\label{fig:imaginaryOnlyKerr}
\end{minipage}
\end{document}